\documentstyle[11pt,gh2001-asp,twoside]{article}
\def\edcomment#1{\iffalse\marginpar{\raggedright\sl#1\/}\else\relax\fi}
\reversemarginpar

\begin{document}
\title{Nested Bars and Associated Morphology of Central Kpc in Disk Galaxies}
 \author{Isaac Shlosman\altaffilmark{1}}
\affil{Joint Institute for Laboratory Astrophysics,
Campus Box 440, University of Colorado, Boulder, CO 80309-0440, USA}

\altaffiltext{1}{JILA Visiting Fellow. Permanent address: Department of
Physics \& Astronomy, University of Kentucky, Lexington, KY 40506-0055}    

\begin{abstract}
We discuss different aspects of nested bar systems, both observational
and theoretical. Such systems consist of a large-scale primary bar leading to
the formation of a sub-kpc size secondary bar, whose pattern speed differs
substantially from that of the main bar. Specifically, we focus on the origin
and gravitational decoupling of nested bars, and their characteristic gas
flows on scales of $\sim 100$~pc$-10$~kpc, with and without gas gravity. 
We find that the gas response in nested bars differs profoundly from that in
single bars, and that no offset dust lanes form in the secondary bars. We also
discuss briefly the importance of nested bar systems for redshifts
corresponding to galaxy formation epoch. 
\end{abstract}

\section{Introduction}

The `barred' branch dominates the Hubble's fork. In the optical band, about
1/3 of all disk galaxies are strongly barred, and an equal fraction is
designated as `intermediate barred' (e.g., Sellwood \& Wilkinson 1993).
In the near-infrared (NIR), the majority of the disks, more than 2/3,
appear to be barred (e.g., Mulchaey \& Regan 1997; Knapen, Shlosman \& Peletier
2000; Grosbol, these proceedings). The importance of bars stems from
their decisive role in galactic dynamics: breaking the axial symmetry of host
galaxies speeds up the dynamical and secular evolution in the disks. The
influence of bars during the galaxy formation epoch, the part they play in
coupling the disks to halo material and the evolution of bar fraction
with redshift only recently have attracted attention.

Theoretical studies of bars have relied on the analysis of dominant families
of stellar orbits  supplemented by 2-D and 3-D
numerical simulations of one- and two-component galactic disks embedded in
`live' or `frozen' halos.  The main effort has been directed towards
understanding the intrinsic bar instability which leads to the radial mass
redistribution as well as triggers and fuels the star formation in the disk,
especially within the central kpc. The presence of {\sl large-scale} bars is
insufficient to maintain the nonstellar (AGN-type) activity within the central
pc, and additional, most probably, dynamical factors are clearly necessary for
this purpose (Shlosman, Begelman \& Shlosman 1990).  

Recent progress in ground-based instruments and the availability of the 
{\it HST\/} has allowed for the first time a meaningful analysis of central
morphology and kinematics. Although our knowledge of the inner regions of disk
galaxies is clearly incomplete, certain patterns in their dynamical evolution
and their relationship to larger and smaller spatial scales have emerged.

At least dynamically, the inner parts of disk galaxies can be defined
by the positions of the inner Lindblad resonances (ILRs), typically
at about 1~kpc from the center. These resonances between the bar pattern speed
and the precession rate of periodic orbits play an important role in filtering
density waves, either stellar or gaseous, propagating between the bar corotation
radius (CR) and the center. They are usually delineated by elevated star
formation rates and the concentration of molecular gas in nuclear rings. These
rings exhibit a rich morphology and can serve as cold gas reservoirs for
fueling the central activity. 

The resolved morphology of central kpc in barred galaxies has revealed
additional grand-design spirals and bars. Defining nested bar systems as
those with more than one bar, stellar or gaseous, we refer to large, kpc-scale
bars as ``primary,'' while the sub-kpc bars as ``secondary.'' The
theoretical rationale behind these definitions is that secondary bars are
believed to form as a result of radial gas inflow along the large-scale bar
and, therefore, are expected to be confined within the ILRs
(Shlosman, Frank, \& Begelman 1989). Below
we review the properties of such nested bars and associated morphologies in
disk galaxies, their dynamical states and the resulting gas flows on scales of
$\sim 100$~pc$-10$~kpc.

\section{Nested Bars: Observations and Initial Statistics}

By a simple analogy with the large-scale bars and their effect on the disk
evolution, secondary bars are expected
to dominate the dynamics inside the central kpc. Although the
sub-kpc bars had been probably detected already by de Vaucouleurs (1974)
and others as optical isophote twists in the central regions of barred
galaxies, they have been interpreted as triaxial bulges. High resolution
ground-based observations have revealed additional small bars
residing within the large-scale stellar bars (Laine et al. 2002 and refs.
therein).

While the first detections of sub-kpc bars were made in stellar light, these
objects can contain arbitrary fractions of gas, and in extreme cases can be
dynamically dominated by molecular gas, as evident from their detection
in CO and the NIR lines of H$_2$ emission (e.g., Ishizuki et al. 1990;
Devereux, Kenney, \& Young 1992; Forbes et al. 1994; Mirabel et al. 1999;
Kotilainen et al. 2000; Maiolino et al. 2000).  CO observations have a rather
low spatial resolution, but do allow the
determination of the offset angle between the large stellar and the
small gaseous bars. It is not yet clear whether stellar- and
gas-dominated sub-kpc bars have a common origin or describe a concurrent
phenomenon. In this section we focus on nested stellar bars, as seen in NIR
and optical starlight. Statistical properties of gaseous bars cannot be
analyzed at present.

Embedded nuclear bars have been detected with the {\it HST\/}
(e.g., Martini \& Pogge 1999; Regan \& Mulchaey 1999). The largest
sample (112) of disk (S0-Sc) galaxies analyzed so far to infer statistical
properties of nested bars in general, and in Seyferts and non-Seyferts in
particular, is that of Laine et al. (2002). The Seyfert sample  consists of
most of the objects in the local universe ($v_{\rm hel}$ $<$ 6000 km~s$^{-1}$)
observed with NICMOS, with the  addition of a few well-known objects. Highly
inclined galaxies have beed removed. But, unless the interaction has been
accompanied by a strong  morphological distortion, interacting galaxies have
not been discarded. The Seyfert sample consisted of 56 objects and was matched
by a control sample of 56 NICMOS non-Seyferts, in
absolute $B$ magnitude, distance, axis ratio and  morphological type.
Ground-based data has been  used for the outer disks. To classify a galaxy as
barred, criteria set in  Knapen et al. (2000) have been applied. Note, that
this strictly conservative approach which was not `curved'  
will lead to lower limits in the bar detection.

Laine et al. (2002) results have confirmed a substantial fraction of
nested bars in disk galaxies, probably in excess of 20--25\%, and that about
1/3 of barred galaxies host second bar. The observed nested bars in
Seyferts and non-Seyferts have revealed an intriguing property --- the
existence of a critical physical length, $l_{crit}\approx$ 1.6~kpc, which
separates the primary and the secondary bars, resulting in a clear bimodal
size distribution with only little overlap (Fig.~1b). When the bar sizes are
normalized to those of the respective host galaxies, $D_{25}$, the overlap
between the two bar species is further reduced and $l_{crit}\approx 0.06$
(Fig.~1a). Both critical sizes appear to be mutually consistent because the
primary  bar lengths exhibit a roughly linear correlation with the parent
galaxy sizes, while the secondary  bar lengths are {\it independent} from the 
sizes of their host galaxies (Fig.~7 in Laine et al.). The importance of this 
result can be inferred from the fact that only in this case the normalized
bar lengths will preserve the identity of both bar groups and there will be no
further mixing between the primary and secondary bars in the normalized size
space. If, for example, both types have had a linear correlation with $D_{25}$
with non-zero slopes, the two bar groups would be separated in physical 
but mixed in the normalized space. 
This is the first time that such a clear separation of primary and secondary
bar lengths has been shown observationally. Because the samples include all the
Hubble types from S0 to Sc, the minimal overlap between the two bar classes
means that this result stands regardless of the morphological class of the
galaxy.  

\begin{figure}[ht!!!!!!]
\vbox to3.85in{\rule{0pt}{3.85in}}
\includegraphics{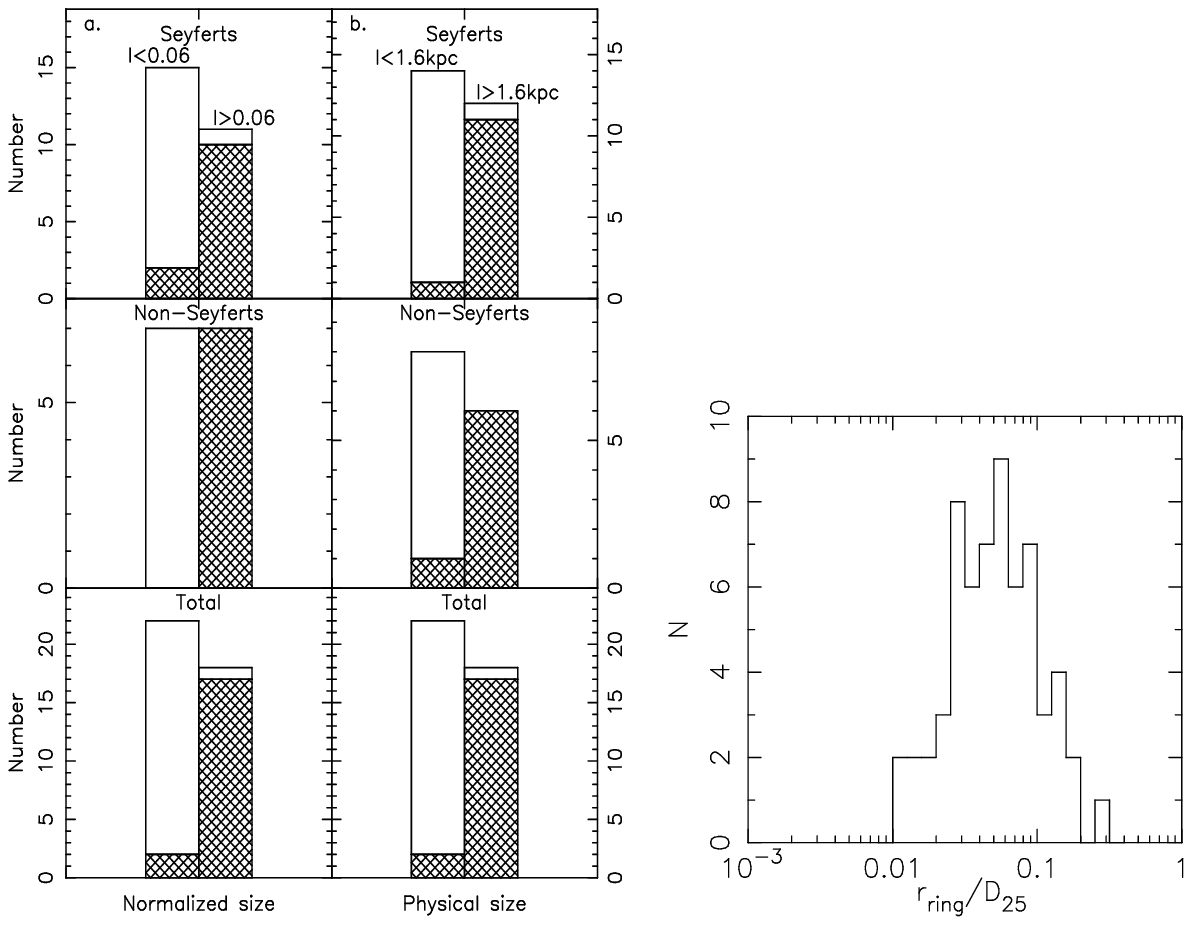}
\parindent 0.8truecm\parbox{4.6in}{\underline{{\it Left:\/} Figure~1.}
Distribution of normalized $(a)$ and  physical $(b)$ primary (cross-hatched)
and secondary (blank) bar sizes. The top panels show Seyferts, the middle
panels non-Seyferts, and the bottom panels display the totals. The bar
lengths, $l$, were normalized by the host galaxy diameter $D_{25}$ and the
resulting values were divided into two groups, $l<$ $l_{\rm crit}=0.06$ and
$l>$ $l_{\rm crit}=0.06$. In physical units $l_{crit}=1.6$~kpc. The lengths
$l_{\rm crit}$ were chosen to minimize the overlap between the two groups.
\underline{{\it Right:\/} Figure~2.} Distribution of normalized nuclear ring
diameters. Data were mostly taken from Buta \& Crocker (1993), with addition
of a few  ``famous'' nuclear ring galaxies (Laine et al. 2002).} 
\end{figure}

A simple explanation for a bimodal distribution of bar sizes 
can be put forward. The linear correlation of the primary bar with
galaxy size means that bars extend to a fixed number of radial
scalelengths in the disk (see also Athanassoula \& Martinet 1980; Martin 1995).
The absence of such a correlation for the secondary bars, together with their
limited range of sizes, hints to a different physical nature of formation and
dynamics compared to the primary bars.                                        
 
Numerical simulations of nested bars show that the secondary bars are
confined to within the ILRs of the primary bars. The outer ILR develops close
to the radius of velocity turnover, or more basically, where the mass
distribution in the inner galaxy switches from 3D to 2D, $\sim$ 1--2~kpc is a
reasonable estimate.  For the early Hubble type galaxies (S0-Sb), this leads
to the appearance of an ILR at about the bulge radius. For the later Hubble
type galaxies  or early types with small bulges, the height of the disk
becomes comparable to the radius of the disk at a point where the 2D disk
approximation breaks down.  

If primary and secondary bars had a similar evolutionary history,
one should expect a linear correlation between the secondary bar length and
$D_{25}$, because of the observed correlations between the disk,
large-scale bar, and bulge sizes. However this correlation is clearly ruled out
by the present data. Large-scale bars are also known to extend just short of
their CR, based on the shapes of their offset dust lanes (Athanassoula 1992).
However, the {\it secondary} bars are not expected to follow this rule or to
possess offset dust lanes (Shlosman 2001; Shlosman \& Heller 2002; Maciejewski
et al. 2002). As discussed in section~3, they must be short of
their CR. This property of secondary bars is expected to destroy any
correlation between their size and that of the parent galaxy.

Laine et al. (2002) have compiled a sample of 62 galaxies with nuclear
rings (mostly from Buta \& Crocker 1993) and determined their normalized size
distribution (Fig.~2). It peaks at $r_{\rm ring}/D_{25}= 0.06$, further
supporting the view that the value 0.06 acts as the dynamical separator between
secondary and primary bars. This result is consistent with the
secondary bars being limited by the size of the ILR, since nuclear rings
are associated with the ILRs.                             

To summarize: $(i)$ nested bars exhibit a bimodal size
distribution with a possibility that ILRs serve as dynamical separators
between primary and secondary bars; $(ii)$ single and primary bars show a
correlation between the ellipticity and the bar length. This result is extended
here for galaxies from S0 to Sc; $(iii)$ only few
{\it single}  bars appear shorter than $l_{crit}=0.06$. Single bars also
have higher average ellipticities than nested bars, and a different
distribution in morphological types; $(iv)$ Seyferts have an excess of bars,
73\% $\pm$ 6\% of Seyferts have at least one bar, against only 50\%
$\pm$ 7\% of non-Seyferts. The statistical significance of this is at
the 2.5$\sigma$ level, and strengthens the result of Knapen et al. (2000) which
was based on smaller samples. Overall it seems that NIR isophote fitting shows 
difficulties when applied to sub-kpc bars, resulting from localized and
distributed sites of dust extinction and bright stars within the central kpc,
leading to a substantial underestimate of bar fraction.

\section{Gas Response in Nested Bar Systems}   

Observational studies of secondary bars customarily assume that their
properties, e.g., gas dynamics and appearance of characteristic
offset dust lanes, are identical to those of large-scale bars (e.g., Martini 
\& Pogge 1999; Regan \& Mulchaey 1999). Theoretically,
gas response to the double bar torquing has been only recently analyzed, when
the pattern speed of the secondary bar, $\Omega_s$ substantially exceeds that
of the primary bar, $\Omega_p$ (Shlosman 2001; Shlosman \& Heller 2002;
Maciejewski et al. 2002). These simulations and general theoretical
considerations support the view that secondary bars are not scaled-down
versions of large bars, and the gas response in nested bars differs
profoundly from that in single bars. The gas flow in  time-dependent
nested-bar potentials is subject to dynamical constraints, such as conditions
minimizing chaos. In order to decrease the fraction of chaotic orbits in the
system, CR of the secondary bar must lie in the vicinity of the
primary-bar ILR (Tagger et al. 1987; Pfenniger \& Norman 1990), constraining
$\Omega_{\rm s}$. Such a dynamical configuration, in principle, poses a
problem for uninterrupted gas inflow towards smaller radii: the gas
flow is repelled from the small bar CR, because of the
rim formed by the effective potential there. Shlosman et al. (1989)
have argued, in essence, that it is the gas self-gravity that overcomes such
repulsion by modifying the underlying potential. In fact, even in the limit of
neglibigle self-gravity in the gas, the flow is capable of crossing the
bar--bar interface, but not in a steady manner and only for a restricted range
of azimuthal angles (Shlosman \& Heller 2002).  

\begin{figure}[ht!!!!!!!!!!!!!!!!!!!!!!!!!]
\vbox to4.96in{\rule{0pt}{4.96in}}
\includegraphics{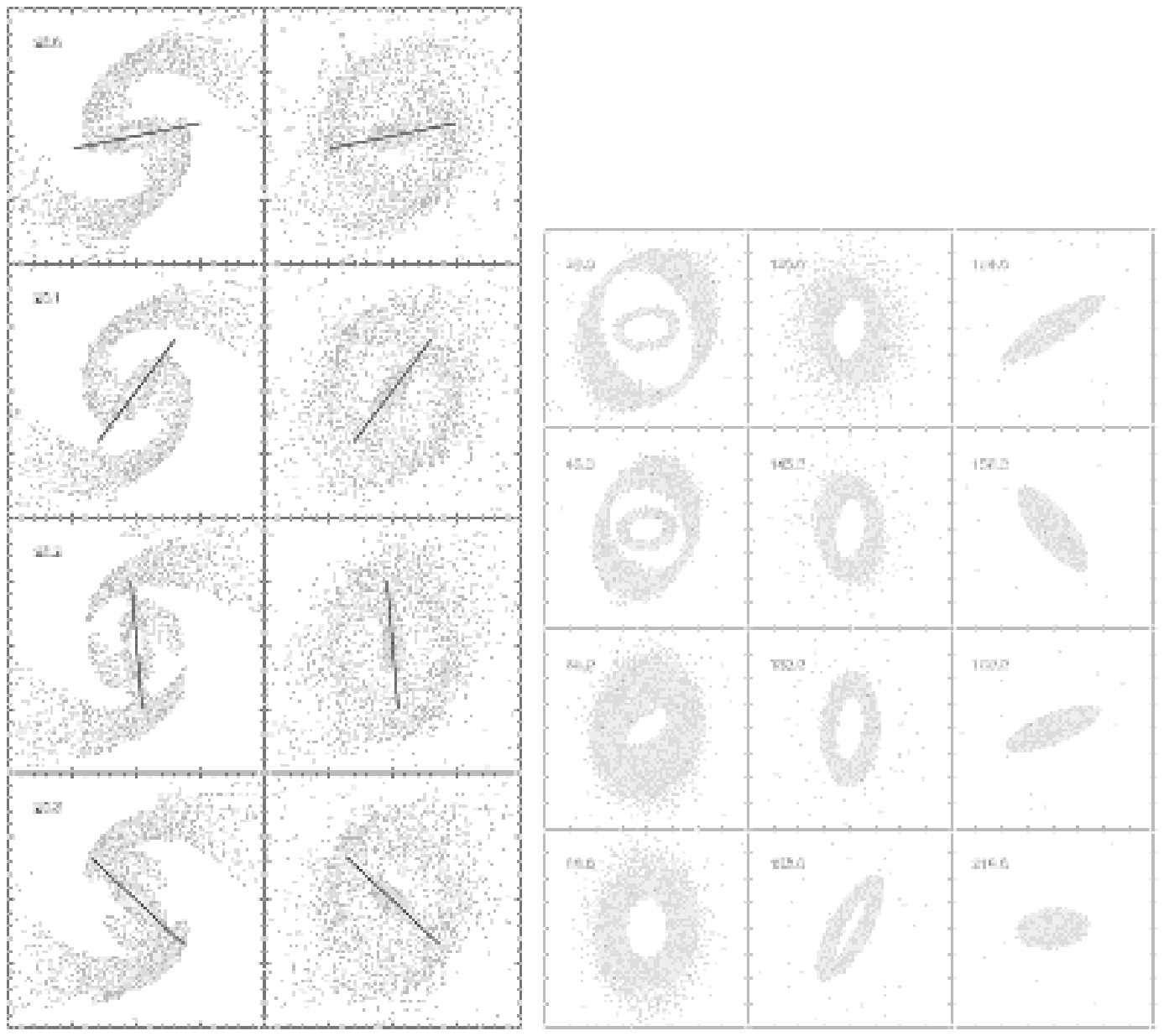} 
\parindent 0.8truecm\parbox{4.7in}{\underline{{\it Left:\/} Figure 3.} Pattern
of shock dissipation (left) and density evolution (right) in the central kpc,
in the frame of reference of the primary bar (horizontal). Positions of the
secondary bar and its length are indicated by a straight line. All rotation is
counter-clockwise. Particles on the left have greater than average
dissipation rate, given by the time derivative of the nonadiabatic
component of internal energy. Note the sharply reduced dissipation in the
innermost secondary bar and ``limb brightening'' enveloping it. Also visible
are two dissipative systems associate with the shocks in the primary bar and
with the trailing shocks in the secondary bar (Shlosman \& Heller 2002).
\underline{{\it Right:\/} Figure 4.} Time evolution of the low-viscosity model:
2D SPH simulation in the background gravitational potential of a barred disk
galaxy (shown face on). The gas response to the bar torquing is displayed in
the primary-bar frame. The primary bar is horizontal and the gas rotation is
counter-clockwise. Note a fast evolution after $t\sim 150$, when the secondary
bar decouples and swings clockwise! The bar is ``captured'' again at $t\sim
211$. Time is given in units of dynamical time. This animation sequence
and others are available in the online edition of Heller et al. (2001).} 
\end{figure}          

To analyze the gas flow in nested bars, one can define the following
three regions: $(i)$ the primary-bar region (outside its ILR), $(ii)$
the bar--bar interface (hereafter the {\it interface}) encompassing the outer
ILR of the large bar and the outer part of the secondary bar, and $(iii)$ the
interior of the secondary bar. In the first region, the gas responds by
forming a pair of large-scale shocks and flows inwards across the interface. 
The flow in the primary bar, outside the interface, is steady and the shock
strength and shape are nearly independent of time. In the second region, 
the flow is time-dependent due to the perturbative effects of the secondary
bar and changing background potential. The inflow proceeds on the
average along the primary-bar minor axis, while an outflow (albeit at a
smaller rate) is directed along its major axis. This happens because the
inflow is driven mainly along the large-scale shocks penetrating the
interface. At the same time the outflow is detected at
angles which  `avoid' the large-scale shocks. The gas, which is repelled by 
the secondary bar is found to enter large-scale shocks while still moving
out, increases the mixing of material with different angular momenta. 
The net effect is an inflow across the CR of the secondary bar (Fig.~3 in
Shlosman \& Heller 2002). The inflow shows
the beat frequency clearly identified with the secondary bar tumbling.
The corresponding mass influx rate is of the order of $0.3\ M_{\rm gas,9}\
M_\odot$ yr$^{-1}$, where $M_{\rm gas,9}$ is the total gas mass in the disk in
units of $10^9\ M_\odot$ within the CR of the large bar. 

Crossing the bar interface, the gas falls towards the third region, the inner
1/2--1/3 of the bar where the flow is very relaxed, with uniform dissipation 
(well below the maximum dissipation in the large-scale shocks), and no
evidence for grand-design shocks. A ``limb brightening'' can be noted at the
edge of the small bar revealing an enhanced density of above-average
dissipating particles outside the oval-shaped central region, which results
from the gas joining the bar from all azimuths. The pattern of shock 
dissipation in nested bars (Fig.~3) allows one to separate the incoming
large-scale shocks from those driven by the secondary bar. Note that two
systems of spiral shocks occur. The shapes of the shocks depend on the angle
between the bars. These spiral shocks driven by the secondary bar may have
observational counterparts. 

A number of factors characterize the gas dynamics in the {\it decoupled}
bars: $(i)$ the time-dependent nature of the gravitational potential;
$(ii)$ the nonsteady gas injection into the secondary bar which proceeds
through the primary shocks penetrating the bar--bar interface.
This phenomenon is absent at the CR of the primary bars. Unstable
orbits in the interface region preclude the secondary bars from extending to
their CR (El-Zant \& Shlosman 2002); $(iii)$ a fast-tumbling secondary bar
which prevents the secondary ILRs from forming. Even
in the case of a long-lived decoupled phase secondary bars are not expected
to slow down. The gas inflow across the interface and the resulting central
concentration can speed-up the bar (Heller, Noguchi, \&
Shlosman 1993, unpublished). The low-Mach-number gas flow is well organized
and capable of following these orbits with little dissipation. Non-linear
orbit analysis reveals that, in the deep interior of the secondary bar, the
$x_1$ orbits have a mild ellipticity and no end-loops. This result is robust.
No offset large-scale shocks form under these conditions.

With no ILRs in a large bar, the offset shocks weaken and recede to its major
axis, becoming ``centered,'' but  do not disappear completely
(Athanassoula 1992). Only two examples of centered shocks have been found 
from more than a hundred bars. During the last decade only one more
potential example has been added to this list (Athanassoula, private
communication). It seems plausible that centered shocks are very rarely
observed because they are so weak.

Knapen et al. (1995) have analyzed the shock  dissipation in a self-consistent
gravitational potential of ``live'' stars and gas {\it before} 
decoupling, when both bars tumble with the same pattern speeds, and when the
gas self-gravity is accounted for. No offset shocks have been found in this
configuration either.                                   

We conclude that no large-scale shocks and consequently no offset dust lanes
will form inside secondary bars, whether they are dynamically coupled
or decoupled. The time-dependent, ``anisotropic'' gas inflow across the
interface found here is a completely new phenomenon inherent to nested
bars. The fate of the gas settling inside secondary bars cannot be decided
without invoking its global self-gravity that will
completely change the nature of the flow. Under the observed
conditions in numerical simulations (gas masses and surface densities) the gas
self-gravity should exert a dominating effect on its evolution (section~4.2).

\section{Nested Bars: Dynamical Decoupling} 
 
Probably the most intriguing property of nested bars is their theoretically
anticipated stage of a decoupling, when each bar exhibits a
different pattern speed (Shlosman et al. 1989). Decoupling is
indirectly supported by the observed random orientation of nested bars.
In principle, the following options exist for secondary bar rotation.
First, both bars corotate, being completely synchronized.
This configuration of nearly orthogonal bars can be a precursor to the future
decoupled phase or continue indefinitely. A simple explanation 
of this phenomenon lies in the existence two main families of periodic
orbits in barred galaxies. The gas responding to the
gravitational torques from the primary bar flows towards the center and
encounters the region of $x_2$ orbits, which it populates. The forming
secondary bar may be further strengthened by the gas gravity, which drags
stars into $x_2$ orbits. The amount of gas accumulating in the ILR
resonance region may be insufficient to cause the dynamical runaway. 

Next, if the secondary bar forms via self-gravitational instability (in
stellar or gaseous disks), it must spin in the direction of the primary bar
with $\Omega_s>\Omega_p$ (Shlosman et al. 1989; Friedli \& Martinet 1993;
Combes 1994; Heller \& Shlosman 1994). The  presence of gas appears to be
imperative for this to occur. Both bars are  dynamically {\sl decoupled} and
the angle between them is arbitrary. Lastly, the secondary bar can rotate in
the opposite sense to the primary bar, resulting from merging (Sellwood \&
Merritt 1994). This appears to be a non-recurrent configuration. 

Although above options make specific predictions verifiable observationally,
the triggering mechanism(s) for decoupling require better
understanding. The computational effort has so far gone into analyzing
self-gravitating systems (e.g., Friedli 1999;
Shlosman 1999). However, dynamical evolution proceeds beyond the formation
of two coupled bars, {\sl even} when gas self-gravity is
neglected (Heller et al. 2001): partial or complete decoupling of a
gaseous bar, depending on the degree of viscosity in the gas, can be
triggered for a prolonged  period of time (section~4.1). Both regimes are
likely to be encountered in nature.                  
\smallskip

\noindent {\it 4.1. Decoupling of Non-Self-Gravitating Bars.\/} The actual
degree of viscosity in the ISM is largely unknown. Heller et al. (2001) have 
investigated the effect of viscosity on the gas settling in the ILR region of
a single large-scale stellar bar. Using identical initial conditions, the 
non-self-gravitating gas was evolved by means of a 2D version of the SPH code
(Heller \& Shlosman 1994), or, alternatively, using the grid code ZEUS-2D, 
exhibiting similar results.

The most spectacular evolution occurred in the low-viscosity model, although
all the models showed similar initial evolution, during which the gas
accumulated in a double ring, corresponding roughly to two
ILRs. In all the models the rings interact hydrodynamically and merge (Fig.~4). 
After merging, a single oval-shaped ring corotates with the primary bar,
leading it by $\phi_{\rm dec}$, the decoupling angle whose value depends on
the gas viscosity. In all the models, the remaining ring becomes increasingly
oval and barlike (Figs.~4, 5), its pattern speed changes abruptly, and it
swings  towards the primary bar, {\sl against} the direction of rotation. In 
the inertial frame, this forming gaseous bar spins in
the same sense as the main bar, albeit with $\Omega_s<\Omega_p$, but in the 
primary-bar frame it tumbles in the retrograde direction!  This is sustained
for about 60 dynamical times, $\sim$ 2--3 $\times
10^9$~yr, until it is captured again by the primary potential. The shape of
the decoupled bar and $\Omega_s$ depend on bar orientation. The eccentricity,
$\epsilon$, reaches a maximum when bars are aligned (Fig.~5). In standard and
high-viscosity models, the secondary bar only librates about the primary bar.
 
The key to understanding this behavior lies in the distribution of gas
particles with Jacobi energy, $E_{\rm J}$. After
merging the ring is positioned close to the energy where the transition from
$x_2$ to $x_1$ (at the inner ILR) occurs. The exact value of this transition
energy is model-dependent, but this is of no importance to the
essence of the decoupling. The crucial difference between the
models comes from $(i)$ the position of the forming gaseous bar on the $E_{\rm
J}$ axis after ring merging and $(ii)$ the value of $\phi_{\rm dec}$.
We note that the role of viscosity here is fundamentally different from that
in nearly axisymmetric potentials (planetary rings, galactic warps, etc.),
where it acts to circularize the orbits.                  
   
As the gaseous bar forms at the angle $\phi_{\rm dec}$ to the primary bar, 
gravitational torques act to align the bars. In the low-viscosity model, 
gaseous bar resides on purely $x_2$ orbits and, therefore, responds to the
torque by speeding up its precession while being pulled backwards,
until it is almost at right angles to the bar potential valley. The
decoupling happens abruptly when $\sim1/2$ of the gas finds itself at $E_{\rm
J}$ below the inner ILR. The absence of $x_2$ orbits at these $E_{\rm J}$
means that the gas loses its stable orientation along the
primary-bar minor axis. The gaseous bar has a much smaller $\epsilon$ in the
fourth quadrant than in the first. Such an asymmetry with respect
to the primary-bar axis ensures that the torques from
the primary bar are smaller in the fourth quadrant. But only for the least
viscous model this becomes crucial, and the torques are
unable to confine the bar oscillation, which continues for a full swing of
$2\pi$. The nuclear bar is trapped again at $\tau\sim 211$, after few
rotations with respect to the large-scale bar. 

\setcounter{figure}{4} 
\begin{figure}[ht!!!!]
\vbox to2.15in{\rule{0pt}{2.15in}}
\includegraphics{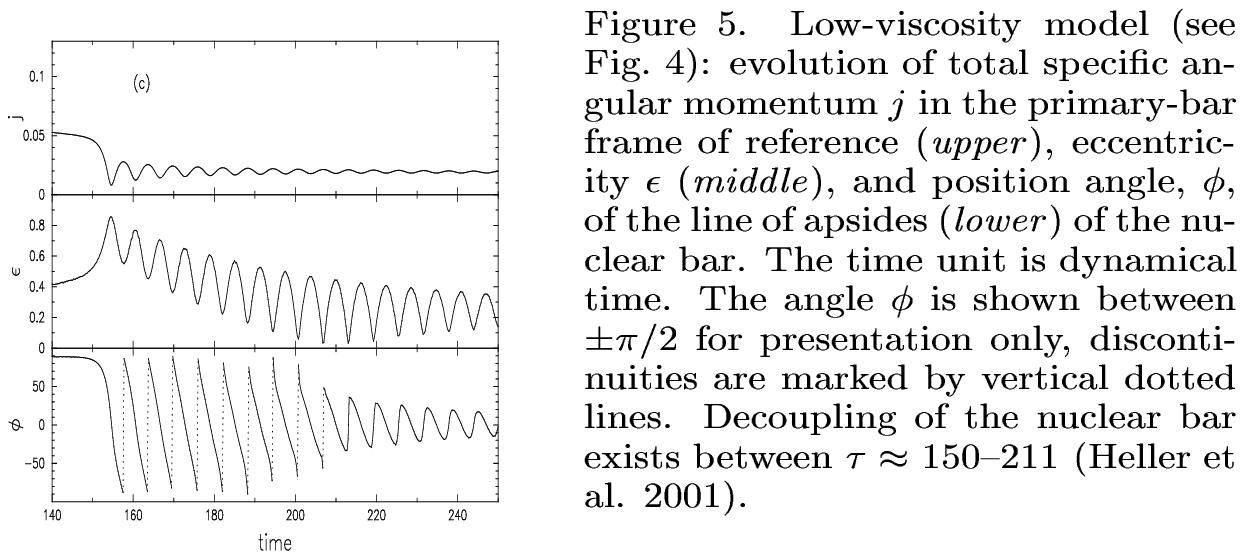}
\end{figure}         

Correlation between $\epsilon$ of the gaseous bar and
its orientation can be tested observationally. Two additional effects should have
observational consequences to decoupling and the periodic
increase in $\epsilon$. First, the gas will cross the inner ILR
on a dynamical timescale. Shlosman et al. (1989) pointed out that  ILRs
present a problem for radial gas inflow because the gas can stagnate there. A
solution was suggested in the form of a global self-gravitational instability
in the nuclear ring or disk, which will generate gravitational torques in the
gas, driving it further in. Recently  Sellwood \& Moore (1999) 
resurrected the idea that ILRs would ``choke'' the gas inflow. However, as we
see here, even non-self-gravitating nuclear rings are prone to dynamical
instability which drives the gas inwards. Moreover, the gas inflow
is expected to be accompanied by star formation along the
molecular bar, having a quasi-periodic, bursting character.      
\smallskip
 
\noindent {\it 4.2. Decoupling of Self-Gravitating Bars.\/} Gas self-gravity
plays a crucial role in the formation and decoupling of 
bars, securing $\Omega_{\rm s} > \Omega_{\rm p}$ (Shlosman et al. 1989;
Shlosman 2001). Simulations which tackle  this issue
have been very limited so far. Friedli \& Martinet (1993, and priv.
communication) and Combes (1994) experienced difficulties in decoupling pure
stellar or gaseous disks. The inner bar in this case is very transient, which
may be a result of an insuffient number of particles.  For a mixed
system the decoupled stage is more prolonged. Simulations explicitly
demonstrate the necessity for gas to be present, in addition to stars, and
confirm that an increased central-mass concentration is important for the
dynamical separation of the outer and inner parts. This can be achieved by
moving the gas along the large-scale bar, accumulating it inside the ILR on
the $x_2$ orbits, modifying the local potential and forming a double ILR. The
gravity of the gas settling into these orbits is sufficient to ``drag'' the
stars along, but such a  configuration still corotates with the primary bar.
The ability of the gas to settle into the $x_2$ orbits depends upon its sound
speed and viscosity. When the gas is too viscous or hot, it will avoid the
ILRs completely and remain on the $x_1$ orbits aligned with the bar (Englmaier
\& Gerhard 1997; Englmaier \& Shlosman 2000; Patsis \& Athanassoula 2000).
Moderately viscous gas will settle into the innermost $x_2$ orbits. The
evolution of nuclear regions in disks, therefore, depends on the (unknown)
equation of state of the ISM. The inclusion of star formation in nuclear rings
has already demonstrated how the resulting increase in viscosity leads to the
mass transfer across the ILRs (Knapen et al. 1995).

Englmaier \& Shlosman (Shlosman 2001) have accounted for
gravitational effects in the gas. Ring evolution was similar to that
described in 4.1, including the swing towards the primary bar. But this
was followed by the rapid growth of {\sl prograde} self-gravitating modes with
$m=2$ and 4 with the pattern speed much larger than $\Omega_{\rm p}$, resulting
in an avalanche-type inflow. A model with the seed 
supermassive BH revealed that inflowing gas feeds the BH at
peak rates, increasing its mass tenfold. A big unknown is
the concurrent star formation, which was neglected in these simulations. Note,
however, that the star formation has low efficiency and can hardly halt this
runaway collapse toward the center.                         

\section{Nested Bars at Higher Redshifts?}

Although galactic disks show little evolution for $z<1$ (Le Fevre et
al. 2000), at higher redhifts the situation is expected to change.
Disk formation and stability should depend on halo shape and its
cuspiness. Analysis of bar dynamics in cuspy axisymmetric halos has
revealed their strong trend to brake (Debattista \& Sellwood 2000) and
the cusp leveling (Weinberg \& Katz 2002). One should question the
very existence of large-scale bars in this environment. Even if the cusps
can be dissolved by the very same processes which form the disks (El-Zant,
Shlosman \& Hoffman 2001), unless the halos are perfectly axisymmetric,
they would impede growth of large bars (El-Zant \& Shlosman 2002).
Hence, first bars may be confined to within the central kpc of growing disks,
at redshifts corresponding to galaxy formation epoch. The
characteristic timescales, dynamical or secular, over which the disks acquire
their present size are not known. One can conjecture that the formation
of disks at large (on $\sim 10$~kpc scale) can be decoupled from dynamics
of central kpc, resulting in bildup of nested bars in the reverse order,
i.e., from smaller to larger spatial scales. The observational verification
of this scenario should probably wait for the {\it NGST\/} and is unrelated to
the reported deficiency of large-scale bars for $z\sim0.5-0.8$ (e.g., Abraham
et al. 1999) which may have an alternative explanation (Jogee et al. 2002).

\section{Conclusions}
 
The lifetime of nested bars is not clear at present and the configuration
itself can be recurrent. The secondary bars can be responsible for accelerating
dynamical and secular evolution within the central kpc, fueling stellar and
nonstellar activity there, and driving the nuclear spiral arms. 
Their role in the formation and initial evolution of galactic disks is far
from being understood. Simulations will clarify the
issues related to nested bar formation and the role of self-gravitating gas
in this process. Coupled with new observational techniques and instruments at
longer wavelengths, they will shed light on the intricate evolution of central
regions within the general context of cosmological evolution.

\acknowledgments I am grateful to my collaborators Amr El-Zant, Peter
Englmaier, Clayton Heller, Shardha Jogee, Johan Knapen, Seppo
Laine and Reynier Peletier, and to my colleagues, too numerous to list here,
for many in-depth discussions on this subject. This work is supported in part
by NASA grants NAG 5-10823, HST GO-08123.01-97A, and WKU-522762-98-6.


\begin{references} 

\reference Abraham. R.G., Merrifield, M.R., Ellis, R.S., Tanvir, N.R.,
     \& Brinchmann, J. 1999, \mnras, 308, 569

\reference Athanassoula, E. 1992, \mnras, 259, 345 

\reference Athanassoula, E., \& Martinet, L. 1980, A\&A, 87, L10

\reference Buta, R., \& Crocker, D.A. 1993, \aj, 105, 1344 

\reference Combes, F. 1994, { Mass-Transfer Induced Activity in Galaxies,}
     ed. I. Shlosman (Cambridge: Cambridge University Press), p. 170   

\reference Debattista, V.P., \& Sellwood, J.A. 2000, \apj, 543, 704

\reference de Vaucouleurs, G. 1974, in IAU Symp. 58, The Formation and
   Dynamics of Galaxies, ed. J.R. Shakeshaft (Dordrecht: Reidel), p. 335

\reference Devereux, N.A., Kenney, J.D.P., \& Young, J.S. 1992, \aj, 103, 784    

\reference El-Zant, A., \& Shlosman, I. 2002, in preparation

\reference El-Zant, A., Shlosman, I., \& Hoffman, Y. 2001, \apj, 760, 636

\reference Englmaier, P., \& Gerhard, O. 1997, \mnras, 287, 57

\reference Englmaier, P., \& Shlosman, I. 2000, \apj, 528, 677  

\reference Forbes, D.A., Kotilainen, J.K., \& Moorwood, A.F.M. 1994, \apj, 433,
    L13

\reference Friedli, D. 1999, Evolution of Galaxies
    on Cosmological Timescales, ed. J.E. Beckman \& T. J.~Mahoney, eds.
    (San Francisco: ASP). p. 88  

\reference Friedli, D., \& Martinet, L. 1993, A\&A, 277, 2  
                            
\reference Heller, C.H., \& Shlosman, I. 1994, \apj, 424, 84 

\reference Heller, C.H., Shlosman, I., \& Englmaier, P. 2001, \apj, 553, 661

\reference Ishizuki, S., Kawabe, R., Ishiguro, M., Okumura, S.K., \& Morita,
   K.-I. 1990, Nat, 344, 224

\reference Jogee, S., Knapen, J.H., Laine, S., Shlosman, I., Scoville, N.Z.,
      \& Englmaier, P. 2002, \apj, submitted

\reference Knapen, J.H., Shlosman, I., \& Peletier, R.F. 2000, \apj, 529, 93 

\reference Knapen, J.H., Beckman, J.E., Heller, C.H., Shlosman, I., \& de 
     Jong, R.S. 1995, \apj, 454, 623  

\reference Kotilainen, J.K., Reunanen, J., Laine, S., \& Ryder, S.D. 2000,
     A\&A, 353, 834 

\reference Laine, S., Shlosman, I., Knapen, J.H., \& Peletier, R.F. 2002, 
     \apj, March 1, in press (astro-ph/0108029) 

\reference Le Fevre, O., et al. 2000, \mnras, 311, 565

\reference Maciejewski, W., Teuben, P.J., Sparke, L.S., \& Stone, J.M.
     2002, \mnras, in press, astro-ph/0109431 

\reference Maiolino, R., Alonso-Herrero, A., Anders, S., Quillen, A., Rieke,
   M.J., Rieke, G.H., \& Tacconi-Garman, L.E. 2000, \apj, 531, 219  

\reference Martin, P. 1995, \aj, 109, 2428 

\reference Martini, P., \& Pogge, R.W. 1999, \aj, 118, 2646   

\reference Mirabel, I.F., et al. 1999, A\&A, 341, 667 

\reference Mulchaey, J.S., \& Regan, M.W. 1997, \apjl, 482, L135

\reference Patsis, P.A., \& Athanassoula, E. 2000, \mnras, 358, 45

\reference Pfenniger, D., \& Norman, C.A. 1990, \apj, 363, 391 

\reference Regan, M.W., \& Mulchaey, J.S. 1999, \aj, 117, 2676

\reference Sellwood, J.A., \& Wilkinson. A. 1993, Rep.Prog.Phys., 56, 173 

\reference Sellwood, J.A., \& Merritt, D. 1994, \apj, 425, 530

\reference Sellwood, J.A., \& Moore, E.M. 1999, \apj, 510, 125    

\reference Shaw, M.A., Combes, F., Axon, D.J., \& Wright, G.S. 1993,
     A\&A, 273, 31    

\reference Shlosman, I. 1999, Evolution of Galaxies 
     on Cosmological Timescales, ed. J.E. Beckman \& T. J.~Mahoney, (San
     Francisco: ASP), p. 100    

\reference Shlosman, I. 2001, The Central Kpc of
     Starbursts \& AGNs, ed. J.H. Knapen et al. (San Francisco: ASP), p.~55  

\reference Shlosman, I., \& Heller, C.H. 2002, \apj, February 1, in press,
           (astro-ph/0109536) 

\reference Shlosman, I., Frank, J., \& Begelman, M.C. 1989, Nat, 338, 45

\reference Shlosman, I. Begelman, M.C., \& Frank, J. 1990, Nat., 345, 679
 
\reference Tagger, M., Sygnet, J.F., Athanassoula, E., \& Pellat, R. 1987,
     \apj, 318, L43 

\reference Weinberg, M.D., \& Katz, N. 2002, preprint, astro-ph/0110632
\end{references}
\end{document}